\documentclass[aps,10pt,prd,notitlepage,showpacs,nofootinbib,superscriptaddress]{revtex4-2}
\usepackage{graphicx}
\usepackage[utf8]{inputenc} 
\usepackage{amsmath}
\usepackage{amssymb}
\usepackage{siunitx}
\usepackage[section]{placeins}
\usepackage{float}
\usepackage{comment}
\usepackage{slashed}
\usepackage[normalem]{ulem}
\usepackage{braket}
\usepackage{comment}

\usepackage{caption}
\usepackage{subcaption}

\usepackage[usenames,dvipsnames]{color}
\usepackage[colorinlistoftodos]{todonotes}
\usepackage[colorlinks=true,citecolor=blue,urlcolor=magenta, pdfborder={0 0 0}]{hyperref}

\usepackage{soul}

\usepackage[commandnameprefix=always]{changes}

\usepackage{slashed}

\usepackage{changes}

\begin{document}
\title{EFT of Non-Markovian $U(1)_X$ Breaking: Dark Matter and Gravitational Waves}

\author{Arnab Chaudhuri}
\email{arnab.chaudhuri@nao.ac.jp }
\affiliation{Division of Science, National Astronomical Observatory of Japan, Mitaka, Tokyo 181-8588, Japan}

\begin{abstract}
We develop an effective field theory (EFT) framework for $U(1)_X$ gauge symmetry breaking in which the dynamics of the order parameter acquire non-local-in-time (``memory'') corrections from a heavy dark sector. Integrating out metastable or slowly equilibrating fields generates temporal kernels in the EFT, yielding a history-dependent effective potential for the $U(1)_X$ scalar. These non-Markovian terms qualitatively alter first-order phase transition dynamics by modifying bubble nucleation, latent heat release, and wall propagation. The resulting stochastic gravitational-wave spectrum exhibits distinctive features such as broadened peaks, asymmetric slopes, and possible secondary ``echoes'' that are absent in conventional Markovian treatments. When the memory-generating sector also participates in dark matter production, the same kernel parameters correlate gravitational-wave signatures with the relic abundance. This work establishes the EFT formalism for non-equilibrium symmetry breaking and highlights testable predictions for upcoming GW observatories and DM searches.\\
\\
\textbf{Keywords:} Non-Markovian effective field theory, Cosmological first-order phase transitions, Gravitational waves and dark matter.
\end{abstract}

\maketitle

\section{Introduction}

Cosmological first-order phase transitions (FOPTs) provide a key connection between extensions of the Standard Model (SM) and observable phenomena such as the relic abundance of dark matter (DM) and stochastic gravitational waves (GWs) \cite{Grojean:2006bp,Caprini:2015zlo,Hogan:1986qda,Domenech:2020ssp,Ellis:2018lca,ATLAS:2018cjd,Durrer:2003ja,Kamionkowski:1993fg,Srivastava:2025oer,Chaudhuri:2025ybh,Chaudhuri:2024vrd,Chaudhuri:2022sis}. In many well-motivated scenarios, an additional $U(1)_X$ gauge symmetry is broken in the early Universe \cite{Langacker:2008yv,Holdom:1985ag,Craig:2015xla,Cirelli:2005uq,Erler:2009jh,Chaudhuri:2025cjp}, generating massive dark gauge bosons, portal interactions with the SM, and potentially stable $U(1)_X$-charged states that can act as DM candidates. If this breaking proceeds through a strong FOPT, it can source a stochastic GW background within reach of future interferometers \cite{Caprini:2015zlo,Caprini:2019egz,Weir:2017wfa,Caprini:2018mtu}.

Most analyses adopt an effective field theory (EFT) description where the dynamics are assumed to be \emph{local in time}: the finite-temperature potential $V_{\rm eff}(\phi,T)$ and kinetic terms depend only on instantaneous field values \cite{Parwani:1991gq,Arnold:1992rz,Dolan:1973qd,Quiros:1999jp}. This Markovian treatment is valid when all relevant sectors equilibrate faster than the characteristic timescales of symmetry breaking. However, dark sectors may contain heavy particles with narrow widths, metastable states, or suppressed couplings. In such cases, equilibration is delayed and significant \emph{memory effects} arise \cite{Boyanovsky:1998aa,Calzetta:1986cq,Anisimov:2008dz,DeSimone:2018efk}.

The Schwinger–Keldysh (SK) real-time finite-temperature formalism provides a natural way to capture these effects: integrating out heavy, slowly equilibrating fields yields non-local-in-time effective actions characterized by memory kernels $K(t-t';T)$, which encode the finite response time of the environment \cite{Berges:2004yj,Prokopec:2003pj,Blaizot:2004bg}. Non-Markovian EFT structures of this kind have been studied in non-equilibrium quantum field theory \cite{Anninos:2008qb,Funakubo:2023cyv}, and our work applies them systematically to cosmological first-order phase transitions.

In this work we construct an EFT framework for $U(1)_X$ breaking that systematically incorporates temporal non-locality. Schematically, the effective action contains terms of the form
\begin{equation}
\mathcal{L}_{\rm EFT} \supset - \Phi^\dagger(t)\!\int_{-\infty}^t\!dt'\,K(t-t';T)\,\Phi(t') + \text{h.c.},
\label{eq:nonlocal}
\end{equation}
where $\Phi$ denotes the scalar responsible for $U(1)_X$ breaking and $K$ is a kernel determined by the underlying spectral density \cite{Moore:2000jw,Arnold:1998cy}. When $K$ is short-ranged, the correction reduces to a local thermal mass shift; by contrast, long-tailed kernels give rise to genuine temporal non-locality.

These memory effects can impact the early-universe dynamics in several ways. First, the bounce equation governing bubble nucleation becomes integro-differential, leading to shifts in the nucleation temperature $T_n$, the inverse duration parameter $\beta/H$, and potentially the bubble wall velocity $v_w$ \cite{Enqvist:1993fm,John:2000zq}. Second, the resulting GW spectrum is modified: non-locality can broaden spectral peaks, alter spectral slopes, and, for oscillatory kernels, induce secondary ``echo'' features in frequency space \cite{Binetruy:2012ze,Caprini:2006jb,Caprini:2009yp}. Finally, if the heavy sector participates in DM production, the delayed response encoded in $K$ modifies the timing of freeze-in or freeze-out and alters chemical potential evolution. This can correlate specific GW spectral imprints with the DM relic abundance \cite{Hall:2009bx,Arcadi:2017kky,Co:2015pka}.

In summary, we provide a systematic derivation of non-local EFT operators for $U(1)_X$ symmetry breaking using SK matching techniques, develop the corresponding framework for bounce dynamics and bubble evolution, and compute the associated GW spectra. We find broadened peaks, modified slopes, and possible echo-like features. Moreover, we demonstrate how these GW signatures correlate with DM abundance, offering a new observational link between early-universe phase transitions and present-day cosmic relics.

\section{Theoretical Framework}

\subsection{$U(1)_X$ Model Setup}

We consider a dark sector with a local $U(1)_X$ gauge symmetry, spontaneously broken by a complex scalar field $\Phi$ of charge $q_X=1$. After symmetry breaking, the associated gauge boson acquires mass
\begin{equation}
m_X = g_X v_X,
\end{equation}
where $g_X$ is the gauge coupling and $v_X$ the vacuum expectation value (VEV) of $\Phi$.

The dark sector Lagrangian, including heavy fermions $\Psi_i$ responsible for memory effects, is
\begin{align}
\mathcal{L}_{\rm DS} &= - \frac{1}{4} X_{\mu\nu} X^{\mu\nu} + |D_\mu \Phi|^2 - V(\Phi) 
+ \sum_i \bar{\Psi}_i (i \slashed{D} - M_i) \Psi_i \nonumber \\
&\quad - \sum_i \big(y_i \Phi \bar{\Psi}_i \Psi_i + {\rm h.c.}\big),
\label{eq:darklag_full}
\end{align}
with scalar potential
\begin{equation}
V(\Phi) = -\mu^2 |\Phi|^2 + \lambda |\Phi|^4.
\end{equation}

The heavy fermions $\Psi_i$, with $M_i \gg v_X$, couple to $\Phi$ via Yukawa-like interactions $y_i$. Integrating out these fermions induces non-local terms in the low-energy EFT of $\Phi$, leading to temporal memory effects in the dynamics of symmetry breaking. Portal couplings to the SM Higgs, such as $\lambda_{\rm portal} |\Phi|^2 |H|^2$, can link the dark and visible sectors but are not required for the derivation of the memory kernel \cite{Arcadi:2017kky,Bian:2018mkl}.

\subsection{Integrating Out the Heavy Sector}

To derive the low-energy EFT, we integrate out the heavy fermions $\Psi_i$ using the Schwinger–Keldysh (SK) real-time formalism at finite temperature \cite{Boyanovsky:1998aa,Prokopec:2003pj}. The generating functional on the forward (1) and backward (2) contours is
\begin{equation}
Z[J_1, J_2] = \int \mathcal{D} \Phi_1 \mathcal{D} \Phi_2 \, \exp \Bigg\{ i \Big[ S_{\rm DS}[\Phi_1, \Psi_1] - S_{\rm DS}[\Phi_2, \Psi_2] + \int d^4 x (J_1 \Phi_1 - J_2 \Phi_2) \Big] \Bigg\}.
\end{equation}

Integrating over $\Psi_i$ yields a one-loop non-local effective action:
\begin{equation}
S_{\rm EFT}[\Phi_1, \Phi_2] = \int d^4x \, \big( |\partial_\mu \Phi|^2 - V(\Phi) \big) - i \sum_i \ln \det \big(i \slashed{D} - M_i - y_i \Phi \big).
\label{eq:SK_eff_action_loop}
\end{equation}

Expanding the logarithm to second order in $y_i$ generates quadratic non-local terms, which in the finite-temperature SK formalism take the form
\begin{equation}
\mathcal{L}_{\rm EFT} \supset - \int_{-\infty}^{t} dt' \, \Phi^\dagger(t) K(t-t';T) \Phi(t').
\label{eq:memory_lagrangian}
\end{equation}

The memory kernel is given by
\begin{equation}
K(t-t';T) = \sum_i y_i^2 \int \frac{d^3 k}{(2\pi)^3} \frac{\sin[\omega_k (t-t')]}{\omega_k} \big[1 - 2 f_F(\omega_k) \big], \quad \omega_k = \sqrt{k^2 + M_i^2},
\label{eq:kernel_explicit}
\end{equation}
with $f_F(\omega_k) = 1/(e^{\omega_k/T}+1)$ the Fermi–Dirac distribution. This kernel encodes the delayed response of the heavy sector, with characteristic timescales set by $1/M_i$ \cite{Calzetta:2008iqa,Berges:2004yj}.

\subsection{Effective Potential with Memory}

The resulting effective potential depends on the history of $\Phi$:
\begin{equation}
V_{\rm eff}[\Phi(t), \{\Phi(t')\}] = V(\Phi(t)) + \int_{-\infty}^t dt' \, \Phi^\dagger(t) K(t-t';T) \Phi(t').
\label{eq:Veff_history}
\end{equation}

In the local (Markovian) limit, $K(t-t') \to \delta(t-t') m_{\rm eff}^2(T)$, reproducing the usual thermal effective potential,
\begin{equation}
V_{\rm eff}^{\rm local}(\Phi,T) = -\mu^2(T) |\Phi|^2 + \lambda |\Phi|^4.
\end{equation}

By contrast, long-tailed kernels lead to integro-differential dynamics:
\begin{equation}
\ddot{\Phi}(t) + 3 H \dot{\Phi}(t) + \frac{\delta V(\Phi(t))}{\delta \Phi^\dagger(t)} + \int_{-\infty}^t dt' K(t-t';T) \Phi(t') = 0.
\label{eq:bounce_memory_full}
\end{equation}

This modification alters bubble nucleation, wall propagation, and phase transition dynamics \cite{Enqvist:1993fm,John:2000zq}. The delayed response can shift the nucleation temperature $T_n$, modify $\beta/H$, and change the effective wall tension, leaving distinct imprints on the GW spectrum \cite{Binetruy:2012ze}. Moreover, if the heavy sector also participates in DM production, the non-locality influences the timing of freeze-out or freeze-in, correlating relic abundance with GW observables \cite{Hall:2009bx,Co:2015pka,Arcadi:2017kky}.

\section{Gravitational Wave Signatures}
\label{sec:GW}

A first-order phase transition generates a stochastic gravitational-wave (GW) background through three main channels: bubble collisions, sound waves, and magnetohydrodynamic (MHD) turbulence in the plasma \cite{Binetruy:2012ze,Caprini:2006jb,Caprini:2009yp,Caprini:2015zlo}. Temporal non-locality modifies each source by smearing nucleation and latent heat release over finite timescales.

\subsection{Bubble Collisions}

Within the envelope approximation, the GW spectrum from bubble collisions is
\begin{equation}
\Omega_{\rm coll}(f) h^2 \simeq 1.67 \times 10^{-5} 
\left(\frac{H_*}{\beta}\right)^2 
\left(\frac{\kappa_{\phi} \alpha}{1+\alpha}\right)^2 
\left(\frac{100}{g_*}\right)^{1/3} 
S_{\rm env}(f),
\label{eq:GW_coll}
\end{equation}
where $H_*$ is the Hubble rate at nucleation, $\beta$ the inverse duration of the transition, $\alpha$ the vacuum-to-radiation energy ratio, $\kappa_\phi$ the fraction of vacuum energy converted to scalar gradients, and $S_{\rm env}(f)$ the spectral shape \cite{Kosowsky:1992vn,Kosowsky:1992rz}.  

Non-locality modifies the nucleation history by effectively reducing $\beta$: 
\begin{equation}
\beta_{\rm mem}^{-1} \sim \frac{d}{dt} \ln \Gamma(T_n)^{-1} 
\simeq \beta^{-1} \left(1 + \gamma \frac{\tau_{\rm mem}}{R_b}\right),
\end{equation}
with $\tau_{\rm mem}$ the kernel memory time, $R_b$ the critical bubble radius, and $\gamma \sim \mathcal{O}(1)$ capturing kernel-shape effects. Larger $\tau_{\rm mem}$ broadens the nucleation-time distribution and can induce secondary structures.

\subsection{Sound Waves and Latent Heat Release}

For subluminal wall velocities $v_w$, the dominant GW signal arises from acoustic waves in the plasma \cite{Hindmarsh:2013xza,Hindmarsh:2015qta}:
\begin{equation}
\Omega_{\rm sw}(f) h^2 \simeq 2.65 \times 10^{-6} 
\left(\frac{H_*}{\beta}\right)
\left(\frac{\kappa_v \alpha}{1+\alpha}\right)^2 
\left(\frac{100}{g_*}\right)^{1/3} 
v_w S_{\rm sw}(f),
\label{eq:GW_sw}
\end{equation}
where $\kappa_v$ denotes the efficiency of latent heat transfer into bulk motion.  

Memory effects delay energy injection into the plasma and modify $\kappa_v$:
\begin{equation}
\kappa_v^{\rm mem} = \kappa_v \, F\!\left(\tfrac{\tau_{\rm mem}}{R_w},\,\tfrac{\beta}{H}\right),
\end{equation}
where $R_w$ is the wall thickness and $F$ a kernel-dependent function. Observable consequences include broadened peaks, skewed spectral slopes, and—in the case of oscillatory kernels—secondary bumps in the spectrum \cite{Binetruy:2012ze,Caprini:2015zlo}.

\subsection{Turbulence Contribution}

MHD turbulence provides a subdominant but non-negligible GW source \cite{Caprini:2009yp,Kahniashvili:2008pf}:
\begin{equation}
\Omega_{\rm turb}(f) h^2 \simeq 3.35 \times 10^{-4} 
\left(\frac{H_*}{\beta}\right)
\left(\frac{\kappa_{\rm turb} \alpha}{1+\alpha}\right)^{3/2} 
\left(\frac{100}{g_*}\right)^{1/3} 
v_w S_{\rm turb}(f),
\label{eq:GW_turb}
\end{equation}
with $\kappa_{\rm turb}$ the fraction of energy in turbulence. Non-local energy deposition reduces the peak amplitude while broadening the profile.

\subsection{Correlation with the Memory Kernel}

The GW peak frequency scales as
\begin{equation}
f_{\rm peak} \sim \frac{\beta_{\rm mem}}{v_w} 
\frac{T_n}{100~{\rm GeV}} 
\left(\frac{g_*}{100}\right)^{1/6},
\end{equation}
so $\beta_{\rm mem}$ and $v_w$ directly map kernel properties into GW observables. Key signatures of non-Markovian dynamics include:
\begin{itemize}
\item Peak broadening: $\Delta f / f_{\rm peak} \sim 1 + \tau_{\rm mem}/R_w$.  
\item Asymmetry: steeper high-frequency falloff relative to the low-frequency slope.  
\item Echoes: oscillatory kernels produce sidebands and secondary peaks \cite{Binetruy:2012ze,Caprini:2015zlo,Caprini:2009yp}.
\end{itemize}

\subsection{Benchmark Points and Phase Transition Parameters}

We illustrate these effects using three benchmark points (BPs) in the $U(1)_X$ parameter space. The microscopic inputs are $\mu$, $\lambda$, $g_X$, the heavy fermion masses $M_i$, and their Yukawa couplings $y_i$. The memory time is estimated as $\tau_{\rm mem}\simeq 1/M_i$, while oscillatory features are characterized by $\omega_{\rm mem}\simeq 2M_i$.  

\begin{table}[h!]
\centering
\begin{tabular}{c|c|c|c|c|c|c|c|c}
\hline
Benchmark & $\mu$ [GeV] & $\lambda$ & $g_X$ & $M_i$ [GeV] & $y_i$ & $\tau_{\rm mem}$ [GeV$^{-1}$] & $\omega_{\rm mem}$ [GeV] & Memory Regime \\
\hline
BP1 & 100 & 0.10 & 0.3 & 300  & 0.5 & $3.3\times 10^{-3}$ & 600  & Weak \\
BP2 & 200 & 0.20 & 0.5 & 600  & 0.7 & $1.7\times 10^{-3}$ & 1200 & Moderate \\
BP3 & 300 & 0.50 & 0.8 & 1000 & 1.0 & $1.0\times 10^{-3}$ & 2000 & Strong/oscillatory \\
\hline
\end{tabular}
\caption{Microscopic benchmark parameters and the corresponding memory scales. $\tau_{\rm mem}\!\sim\!1/M_i$ sets the correlation time, while $\omega_{\rm mem}\!\sim\!2M_i$ controls echo spacing in frequency space.}
\label{tab:benchmarks}
\end{table}

The corresponding macroscopic phase transition parameters are summarized in Table~\ref{tab:alpha_beta_Tn_benchmarks}. These values are obtained from bounce action calculations with memory kernels and are representative of the weak, moderate, and strong regimes.  

\begin{table}[ht]
\centering
\begin{tabular}{c c c c c}
\hline\hline
Label (BP) & $\alpha$ & $\beta/H$ & $T_n$ [GeV] & Memory strength \\
\hline
1 & $1\times 10^{-2}$ & $300$ & $90$  & Weak--moderate \\
2 & $6\times 10^{-2}$ & $120$ & $160$ & Moderate \\
3 & $4\times 10^{-1}$ & $40$  & $280$ & Strong/oscillatory \\
\hline\hline
\end{tabular}
\caption{Macroscopic phase transition parameters corresponding to the benchmarks in Table~\ref{tab:benchmarks}.}
\label{tab:alpha_beta_Tn_benchmarks}
\end{table}
Since memory effects also shift dark matter freeze-out/freeze-in timing, the benchmark dependence in Tables~\ref{tab:benchmarks} and \ref{tab:alpha_beta_Tn_benchmarks} simultaneously controls both GW features and relic abundance, strengthening the observational complementarity.

\begin{figure}[h!]
\centering
\includegraphics[width=0.7\textwidth]{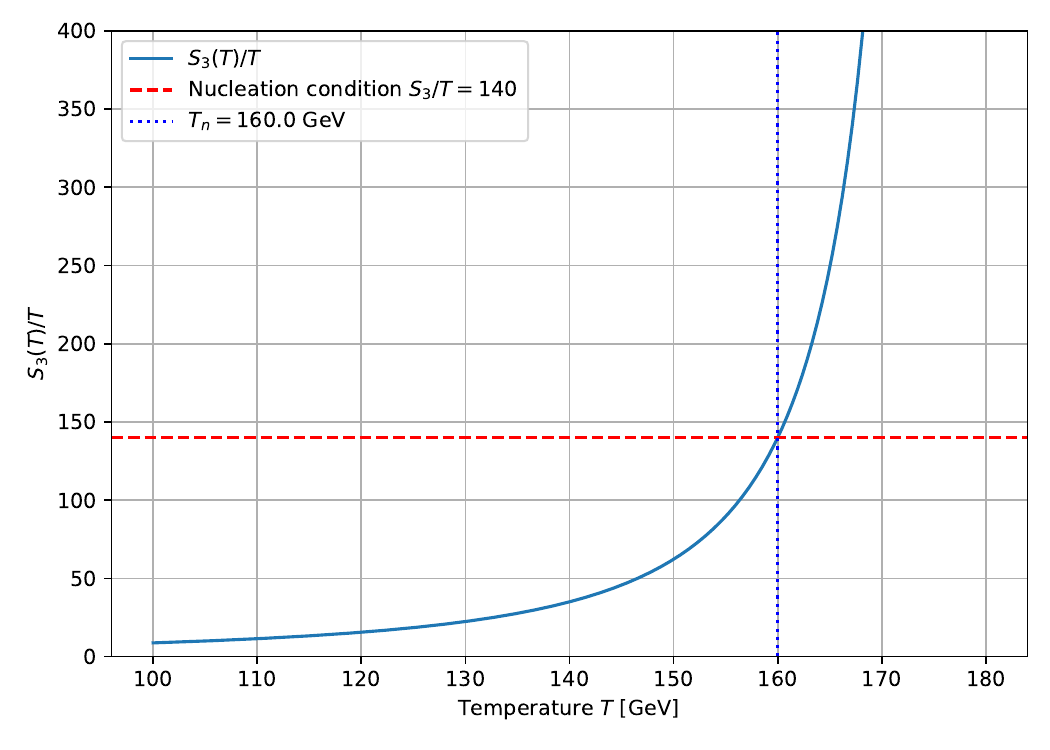}
\caption{Bounce action $S_3(T)/T$ versus temperature for BP2. 
The nucleation temperature $T_n$ corresponds to the point where $S_3(T)/T$ crosses the critical value of $\sim 140$, shown by the dashed red line.}
\label{fig:bounce_action}
\end{figure}

As an illustration, Fig.~\ref{fig:bounce_action} shows the bounce action $S_3(T)/T$ as a function of temperature for BP2, computed using the \texttt{CosmoTransitions} package \cite{Wainwright:2011kj}. 
The nucleation temperature $T_n$ is defined by the point at which $S_3(T)/T \simeq 140$, marked by the intersection of the curve with the horizontal dashed line. This demonstrates explicitly how the quoted $T_n$ values are obtained in our analysis.

\subsection{Echo Modelling}

Oscillatory kernels of the form
\begin{equation}
K(t-t';T) \supset \kappa_{\rm osc} e^{-(t-t')/\tau_{\rm mem}} 
\cos\!\big[\omega_{\rm mem}(t-t')+\varphi\big],
\label{eq:kernel_osc}
\end{equation}
introduce sidebands in frequency space with spacing
\begin{equation}
\Delta f_{\rm echo} \simeq \frac{M_i}{\pi},
\label{eq:omega_mem_micro}
\end{equation}
and amplitude $\kappa_{\rm osc}\!\sim y_i^2/M_i^2$.  

The resulting GW spectrum can be modeled as
\begin{equation}
\Omega_{\rm GW}^{\rm tot}(f) = \Omega_{\rm GW}^{(0)}(f) 
+ \sum_{n=1}^{N_{\rm echo}} A_n \, 
\Omega_{\rm GW}^{(0)}(f - n\Delta f_{\rm echo}) 
\, e^{-\gamma_n (f-n\Delta f_{\rm echo})^2},
\label{eq:GW_echo}
\end{equation}
where $\Omega_{\rm GW}^{(0)}$ is the baseline spectrum. In the sinusoidal limit this simplifies to
\begin{equation}
\Omega_{\rm GW}^{\rm tot}(f) \simeq \Omega_{\rm GW}^{(0)}(f) 
\left[ 1 + \epsilon \cos\!\left( 2\pi \frac{f}{\Delta f_{\rm echo}} + \phi \right) \right].
\label{eq:GW_echo_sinus}
\end{equation}

\begin{figure}[h!]
\centering
\includegraphics[width=0.75\textwidth]{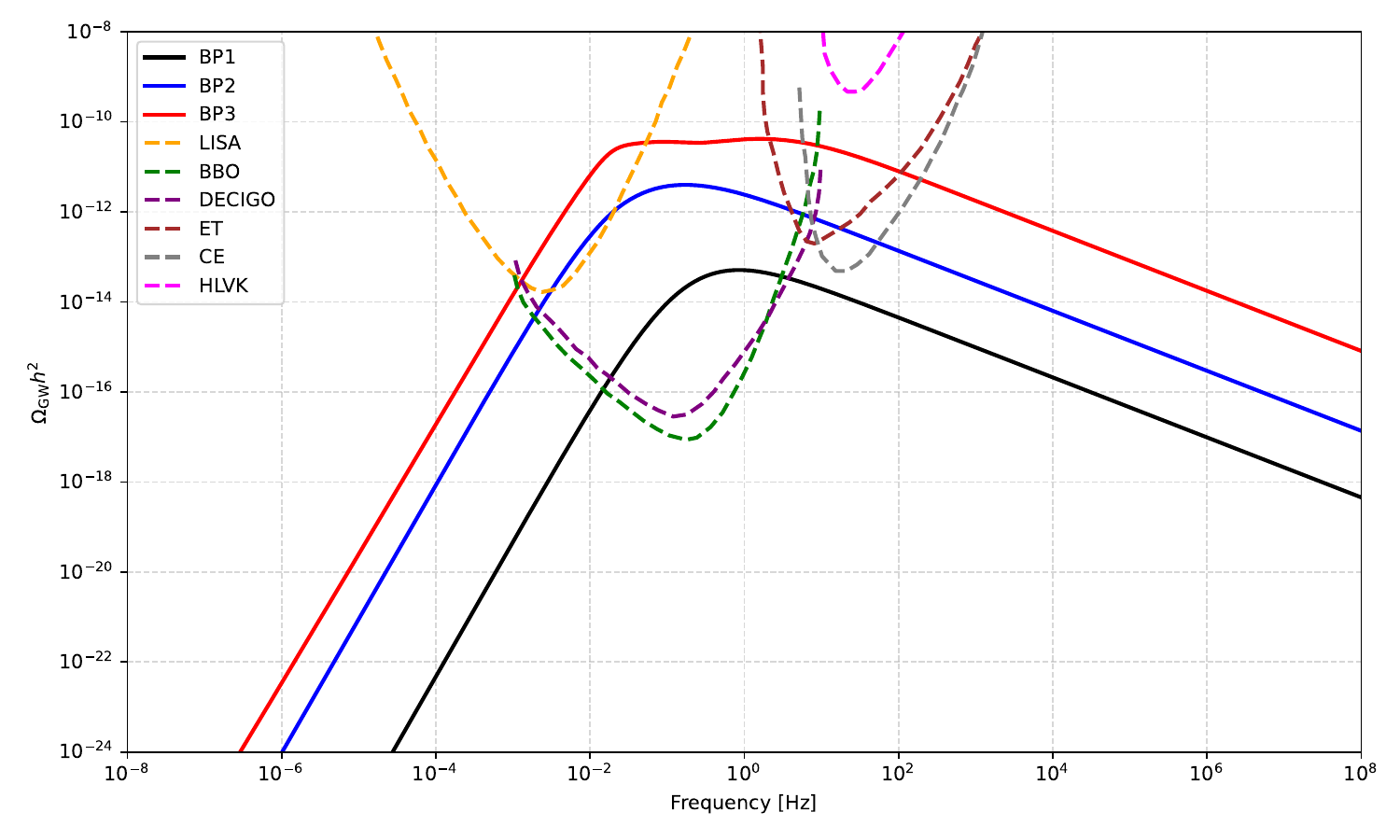}
\caption{Gravitational-wave spectra for BP1--BP3. BP3 exhibits echo features with spacing $\Delta f_{\rm echo}$ from Eq.~\eqref{eq:omega_mem_micro}. Sensitivity curves for LISA, BBO, DECIGO, ET, CE, and HLVK are overlaid.}
\label{fig:GW}
\end{figure}
These echo features, if present in the mHz–Hz frequency band, lie within the sensitivity reach of LISA, DECIGO, and BBO, providing a direct observational test of non-Markovian dynamics.

\section{Dark Matter Phenomenology}
\label{sec:DM}

The dark $U(1)_X$ sector with scalar $\Phi$ and gauge boson $X_\mu$ naturally admits viable dark matter (DM) candidates. However, once memory effects are included, the dynamics of relic production are qualitatively altered: the non-local kernel $K(t-t')$ modifies both the thermal effective potential and the freeze-in/freeze-out history.

\subsection{Relic Scenarios}

Two broad scenarios arise:

\paragraph{Stable relic from the heavy sector.}
If the dark gauge boson $X_\mu$ or an additional $U(1)_X$-charged fermion $\chi$ is stabilized by a remnant $\mathbb{Z}_2$, the relic density is determined by thermal freeze-out or freeze-in. Interactions with the Standard Model (SM) proceed via the Higgs portal,
\begin{equation}
\mathcal{L}_{\rm portal} = \kappa |H|^2 |\Phi|^2 ,
\end{equation}
and possible kinetic mixing,
\begin{equation}
\mathcal{L}_{\rm kin.mix} = - \frac{\epsilon}{2} F^{Y}_{\mu\nu} F^{X\,\mu\nu}.
\end{equation}
In the Markovian limit, the abundance follows from the standard Boltzmann equation. With a memory kernel $\mathcal{K}(t-t')$, however, the Boltzmann equation becomes an integro-differential form,
\begin{equation}
\frac{d n_\chi}{dt} + 3Hn_\chi = -\int^t dt'\, \mathcal{K}(t-t')\, \langle\sigma v\rangle \, \big[n_\chi^2(t') - n_{\chi,{\rm eq}}^2(t')\big],
\label{eq:nonlocalBoltz}
\end{equation}
which delays freeze-out and alters the relic yield.

\paragraph{Asymmetric dark matter.}
Alternatively, if the phase transition generates a particle–antiparticle asymmetry (e.g. through complex $\Phi$-dependent Yukawa couplings), then memory effects in bubble dynamics play a role analogous to electroweak baryogenesis. The CP-violating source acquires a nonlocal structure,
\begin{equation}
S_{\rm CP}(t) = \int^t dt'\, K_{\rm CP}(t-t')\, \partial_\mu \theta(t') J^\mu ,
\end{equation}
with $\theta$ the phase of $\Phi$. The resulting asymmetry, and hence relic abundance, directly depends on the kernel properties.

\subsection{Relic Density Mapping}

It is convenient to define an \emph{effective annihilation rate},
\begin{equation}
\langle \sigma v \rangle_{\rm eff} \equiv \int_0^\infty d\tau \, K(\tau)\, e^{-\Gamma \tau}\, \langle\sigma v\rangle ,
\end{equation}
where $\Gamma$ denotes the relaxation rate of the bath.  
For freeze-out, the relic density can be approximated as
\begin{equation}
\Omega_{\chi} h^2 \simeq \frac{1.07\times 10^9 \,{\rm GeV}^{-1}}{M_{\rm Pl}} \, \frac{x_f}{\sqrt{g_*}} \, \frac{1}{\langle \sigma v \rangle_{\rm eff}},
\end{equation}
with $x_f = m_\chi/T_f$.  
For freeze-in, $\Omega_\chi$ is instead obtained from integrating the production rate and is proportional to $\langle\sigma v\rangle_{\rm eff}$ rather than its inverse. For freeze-in production, the relic density instead scales with the integral of the production rate, 
\begin{equation}
\Omega_\chi h^2 \;\simeq\; \frac{1.09 \times 10^{27}}{g_*^{3/2} M_{\rm Pl}} 
\int_{T_{\rm min}}^{T_{\rm max}} \frac{dT}{T^6} \,
\langle \sigma v \rangle_{\rm eff}(T) ,
\end{equation}
where $T_{\rm max}$ is set by the reheating temperature and $T_{\rm min}$ by the point where production becomes negligible. 
In this case, $\Omega_\chi$ grows linearly with $\langle\sigma v\rangle_{\rm eff}$, in contrast to the inverse scaling of the freeze-out scenario \cite{Hall:2009bx,Arcadi:2017kky}.

\begin{figure}[t]
    \centering
    \includegraphics[width=0.7\textwidth]{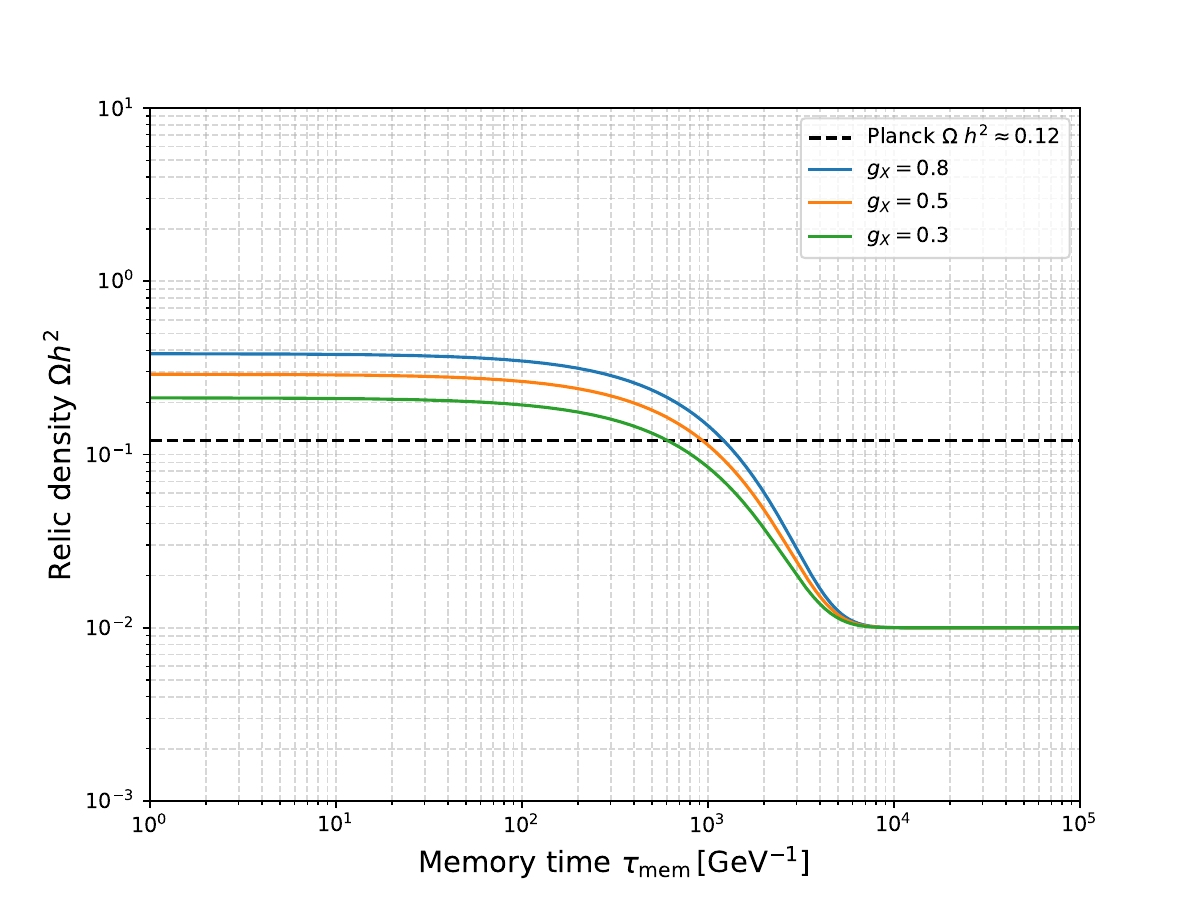}
    \caption{Relic density $\Omega_\chi h^2$ versus $\tau_{\rm mem}$ for $g_X=0.1,0.3,0.5$. Larger $\tau_{\rm mem}$ reduces annihilation efficiency and raises the relic density. The horizontal band shows the Planck value $\Omega_\chi h^2 \approx 0.12$.}
    \label{fig:relic_vs_tau}
\end{figure}

Figure~\ref{fig:relic_vs_tau} shows the relic density as a function of $\tau_{\rm mem}$ for fixed $g_X$. Small $\tau_{\rm mem}$ reproduces the Markovian limit, while larger values suppress annihilation and drive overproduction. The horizontal band marks the Planck value $\Omega_\chi h^2 \simeq 0.12$~\cite{Planck:2018vyg}, selecting a narrow allowed range of $(g_X, \tau_{\rm mem})$. This provides a direct cosmological constraint on the memory timescale.  

We do not explicitly display the $m_\chi$ dependence here, since it follows the standard freeze-out scaling $\Omega_\chi h^2 \propto m_\chi/\langle\sigma v\rangle_{\rm eff}$ in the Markovian limit. The role of the memory kernel is simply to rescale the effective annihilation cross section, so that increasing $\tau_{\rm mem}$ shifts the entire $\Omega_\chi(m_\chi)$ curve upward without altering its qualitative shape. Hence, the relic-density constraints in Fig.~\ref{fig:relic_vs_tau} capture the combined effect of mass and memory, and a separate $m_\chi$ plot would not add new information.

Finally, since the same kernel parameters also control bubble nucleation and GW spectra, the relic density constraints are directly correlated with the GW signatures discussed in Sec.~\ref{sec:GW}. This establishes a unified observational test of memory effects in the dark sector.


\section{Discussion, Outlook and Conclusion}

We have developed an effective field theory framework for symmetry breaking in a hidden $U(1)_X$ sector where time-nonlocal dynamics are systematically incorporated. Using the Schwinger–Keldysh formalism, we derived memory kernels that enter both the effective potential and the transport equations. This represents a conceptual departure from conventional EFT treatments, which implicitly assume Markovianity, and it allows us to capture qualitatively new dynamics in the early universe. 

Our results demonstrate that memory effects reshape the dynamics of first-order phase transitions. Bubble nucleation is governed by integro-differential bounce equations, leading to shifts in the nucleation temperature $T_n$ and the parameter $\beta/H$, while wall propagation exhibits history-dependent overshoot and undershoot phenomena. These modifications feed directly into the stochastic gravitational-wave background. As shown in Fig.~\ref{fig:GW}, the GW spectra for benchmark points BP1–BP3 exhibit broadened peaks and tilted slopes; only in the strong-memory benchmark BP3 do clear echo-like secondary features emerge. The echo spacing $\Delta f_{\rm echo}$ follows directly from the microphysical kernel [Eq.~\eqref{eq:omega_mem_micro}], providing a sharp theoretical prediction that is absent in conventional treatments. Importantly, the predicted signals fall into the sensitivity ranges of LISA, DECIGO, BBO, ET, CE, and HLVK, underscoring the observational relevance of this framework.

On the dark-matter side, the same kernel parameters control the effective annihilation rate. The relic density grows as the memory timescale $\tau_{\rm mem}$ increases, reflecting the suppression of annihilation efficiency. Figure~\ref{fig:relic_vs_tau} illustrates this effect for representative couplings $g_X$: while short-memory behavior reproduces the standard result, large $\tau_{\rm mem}$ drives $\Omega_\chi h^2$ above the Planck-preferred band $\Omega_\chi h^2 \simeq 0.12$~\cite{Planck:2018vyg}. This selects only narrow regions in $(g_X,\tau_{\rm mem})$ as viable, showing that cosmological relic density measurements provide a direct constraint on nonlocal dark-sector dynamics.

Taken together, Figs.~\ref{fig:GW} and \ref{fig:relic_vs_tau} highlight a central outcome of this work: the same memory kernel simultaneously controls the GW spectrum and the DM relic abundance. This establishes a dual observational handle on the hidden sector. In particular, the presence of echo-like features in GW data would provide strong evidence for long memory timescales, which, as Fig.~\ref{fig:relic_vs_tau} shows, have direct consequences for the allowed dark matter parameter space. The complementarity between GW interferometers and cosmological relic-density measurements offers a synergistic path to probing hidden-sector dynamics.

The conceptual novelty of this work lies in extending EFT methods to consistently include temporal nonlocality in gauge theories. While memory effects are generically expected whenever slowly relaxing heavy fields couple to an order parameter, they have been neglected in almost all cosmological studies. Our analysis shows that such effects are not only tractable but also phenomenologically decisive, producing correlated signatures in gravitational waves and dark matter.

Looking forward, several directions warrant further exploration. Nonperturbative lattice studies of memory-loaded phase transitions would provide benchmarks for bounce actions and wall dynamics. Generalizations to non-Abelian sectors or multi-field systems could reveal richer kernel structures and new observable patterns. The interplay with baryogenesis is especially intriguing, as memory-enhanced CP-violating sources could open new mechanisms for generating the baryon asymmetry, again tied to GW and DM observables. Finally, combining the projected sensitivities of LISA, DECIGO, BBO, and next-generation terrestrial interferometers with precision cosmology offers the prospect of significantly constraining the parameter space where non-Markovian new physics is testable.

In conclusion, memory-loaded $U(1)_X$ breaking constitutes a qualitatively new paradigm for early-universe cosmology. The dual correlation between dark matter relic density and gravitational-wave spectra, exemplified by Figs.~\ref{fig:GW} and \ref{fig:relic_vs_tau}, establishes time-nonlocality as a distinctive and potentially observable imprint of hidden-sector dynamics. This framework paves the way for a new generation of studies in which memory effects are elevated from a theoretical curiosity to a practical probe of physics beyond the Standard Model.

\section{Acknowledgment}
The work of AC was supported by the Japan Society for the Promotion of Science (JSPS) as a part of the JSPS Postdoctoral Program (Standard), grant number JP23KF0289.

\appendix
\section{Derivation of the Memory Kernel in the EFT}
\label{app:kernel}

In this appendix we derive the nonlocal-in-time operators that appear in the effective field theory after integrating out heavy fields. We work in the real-time Schwinger–Keldysh (SK) formalism at finite temperature and present both an analytic sketch of the derivation and numerical evaluations of representative kernels.

\subsection{Setup and SK matching}

Consider a heavy Dirac fermion $\chi$ of mass $M$ (carrying $U(1)_X$ charge) coupled to the order parameter $\Phi$ via the Yukawa interaction
\begin{equation}
\mathcal{L}_{\rm int} = - y \, \Phi \, \bar{\chi}_L \chi_R + \text{h.c.}.
\label{eq:yukawa_app}
\end{equation}
The SK generating functional on the closed-time contour $\mathcal{C}$ yields, after integrating out $\chi$,
\begin{equation}
S_{\rm eff}[\Phi] = S_{\Phi} - i \, \mathrm{Tr}_{\mathcal{C}} \log \big[i\slashed{\partial} - M - y \Phi \big].
\end{equation}
Expanding the logarithm to second order in $y$ generates bilinear nonlocal terms of the form
\begin{equation}
\Delta S_{\rm eff} \supset - \int_{\mathcal{C}} d^4x \, d^4x' \,
\Phi^\dagger(x)\, \Pi(x-x')\, \Phi(x'),
\label{eq:nonlocal_action_app}
\end{equation}
with $\Pi$ the fermion polarization evaluated on the contour. The retarded component is
\begin{equation}
\Pi^R(t,\mathbf{x}) = i\theta(t)\langle [\mathcal{O}(t,\mathbf{x}),\mathcal{O}^\dagger(0)]\rangle_T,
\qquad \mathcal{O}=y\,\bar\chi_L\chi_R.
\end{equation}

\subsection{Analytic structure and physical limits}

In momentum space the retarded self-energy can be written (schematically) as
\begin{equation}
\Pi^R(\omega,\mathbf{k}) = y^2 \int \frac{d^3p}{(2\pi)^3}\,
\frac{f_F(E_{\mathbf p})-f_F(E_{\mathbf p+k})}{\omega + E_{\mathbf p} - E_{\mathbf p+k} + i0^+},
\end{equation}
where $f_F(E)=1/(e^{E/T}+1)$. Two regimes are particularly important:

\begin{itemize}
\item \textbf{Thermal / light regime} ($M \lesssim T$): thermal occupation numbers are \(\mathcal{O}(1)\), the dissipative (imaginary) part of $\Pi^R$ is unsuppressed, and the non-local kernel can have sizeable amplitude on timescales $\tau\sim 1/T$ or longer.
\item \textbf{Heavy regime} ($M \gg T$): the on-shell processes that provide an imaginary part are Boltzmann-suppressed. At leading order one finds a schematic expansion
\begin{equation}
\Pi^R(\omega) \simeq \frac{y^2}{16\pi^2}\omega^2 \ln\!\left(\frac{M^2}{\mu^2}\right)
+ i\, y^2\,\omega\, e^{-M/T} + \cdots,
\label{eq:Pi_heavy}
\end{equation}
so that dissipative / memory amplitudes are suppressed by $\sim e^{-M/T}$. The real part produces local operators (mass and wavefunction renormalization) and, to leading power, yields local corrections in the EFT.
\end{itemize}

\noindent\textbf{Boxed conclusion (physical-regime tension).} If the heavy states satisfy $M/T\gg 1$ and have widths $\Gamma \sim \mathcal{O}(M)$, then the imaginary (dissipative) part of $\Pi^R$ — and therefore the amplitude of genuinely non-Markovian memory kernels — is exponentially suppressed $\propto e^{-M/T}$. Significant memory effects therefore require either (i) $M \lesssim \mathcal{O}(\text{a few})\times T$ so thermal excitations are not Boltzmann-suppressed, or (ii) the presence of a meta-stable mediator (small decay width $\Gamma \ll M$) whose pole structure produces long-lived, oscillatory kernels. The numerical examples below illustrate both possibilities.

\subsection{Time-domain kernels: numerical evaluation}

\begin{figure}[h!]
\centering
\includegraphics[width=0.5\textwidth]{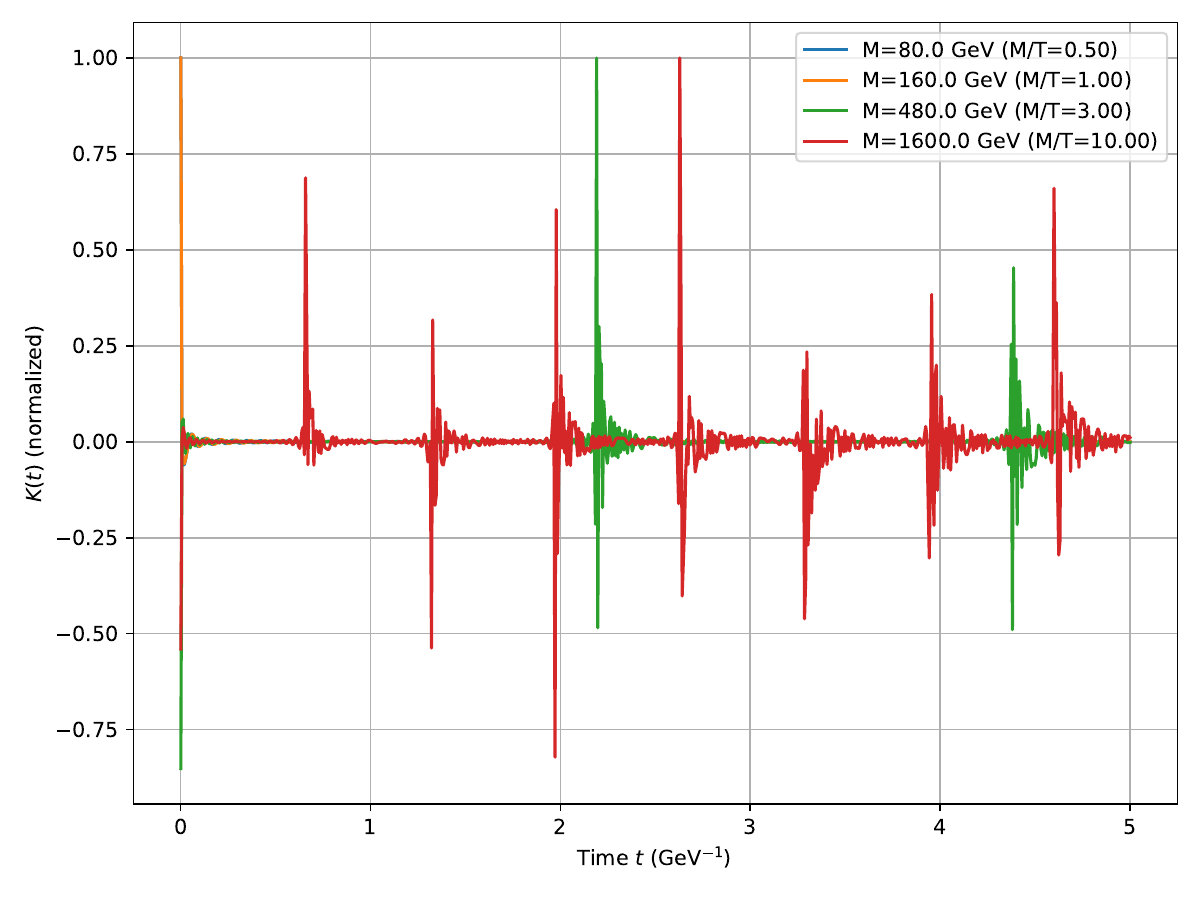}
\caption{Normalized fermion-loop memory kernel $K(t)$ for several $M/T$ values (example: $T=160$~GeV, $y=0.7$). Curves are normalized to unity to highlight the change in temporal shape as $M/T$ varies.}
\label{fig:fermion_kernel_norm}
\end{figure}

\begin{figure}[h!]
\centering
\includegraphics[width=0.5\textwidth]{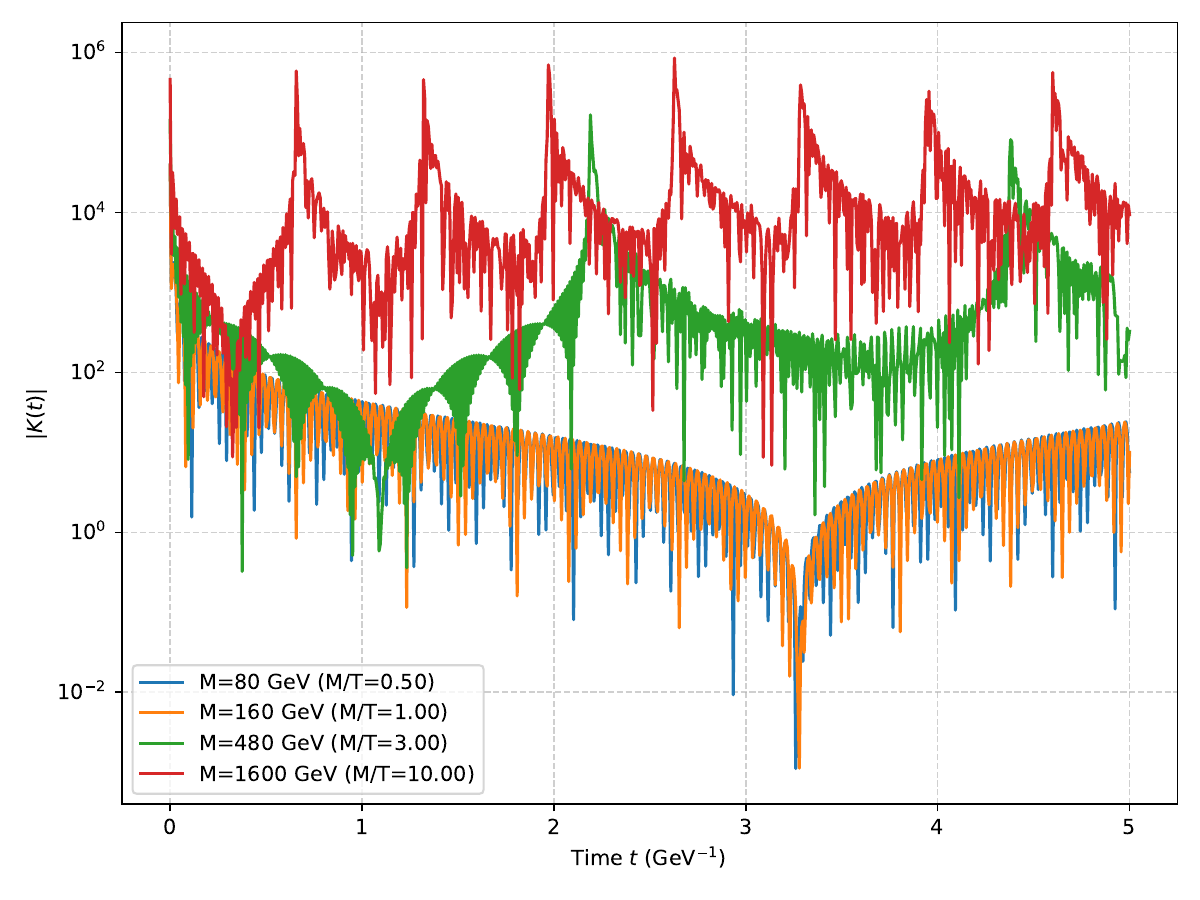}
\caption{Absolute kernel amplitude $|K(t)|$ (log scale). For large $M/T$ the amplitude is strongly suppressed, consistent with the Boltzmann factor $e^{-M/T}$ in Eq.~\eqref{eq:Pi_heavy}.}
\label{fig:fermion_kernel_abs}
\end{figure}

A convenient representation for the time-domain kernel used in the main text is
\begin{equation}
K(t) \;=\; y^2 \int \frac{d^3k}{(2\pi)^3}\,\frac{\sin(\omega_k t)}{\omega_k}\,[1-2f_F(\omega_k)],
\qquad \omega_k=\sqrt{k^2+M^2},
\label{eq:K_numeric}
\end{equation}
which follows from the retarded self-energy and its Fourier transform in the SK formalism. We evaluated Eq.~\eqref{eq:K_numeric} numerically for representative values of $M/T$ and $y$ (details of the numerical integration and convergence checks are given below). Two representative outputs are displayed in Figs.~\ref{fig:fermion_kernel_norm}--\ref{fig:fermion_kernel_abs}.

Figure~\ref{fig:scalar_kernel} shows a toy damped-oscillatory kernel relevant for a meta-stable bosonic mediator:

\begin{figure}[h!]
\centering
\includegraphics[width=0.5\textwidth]{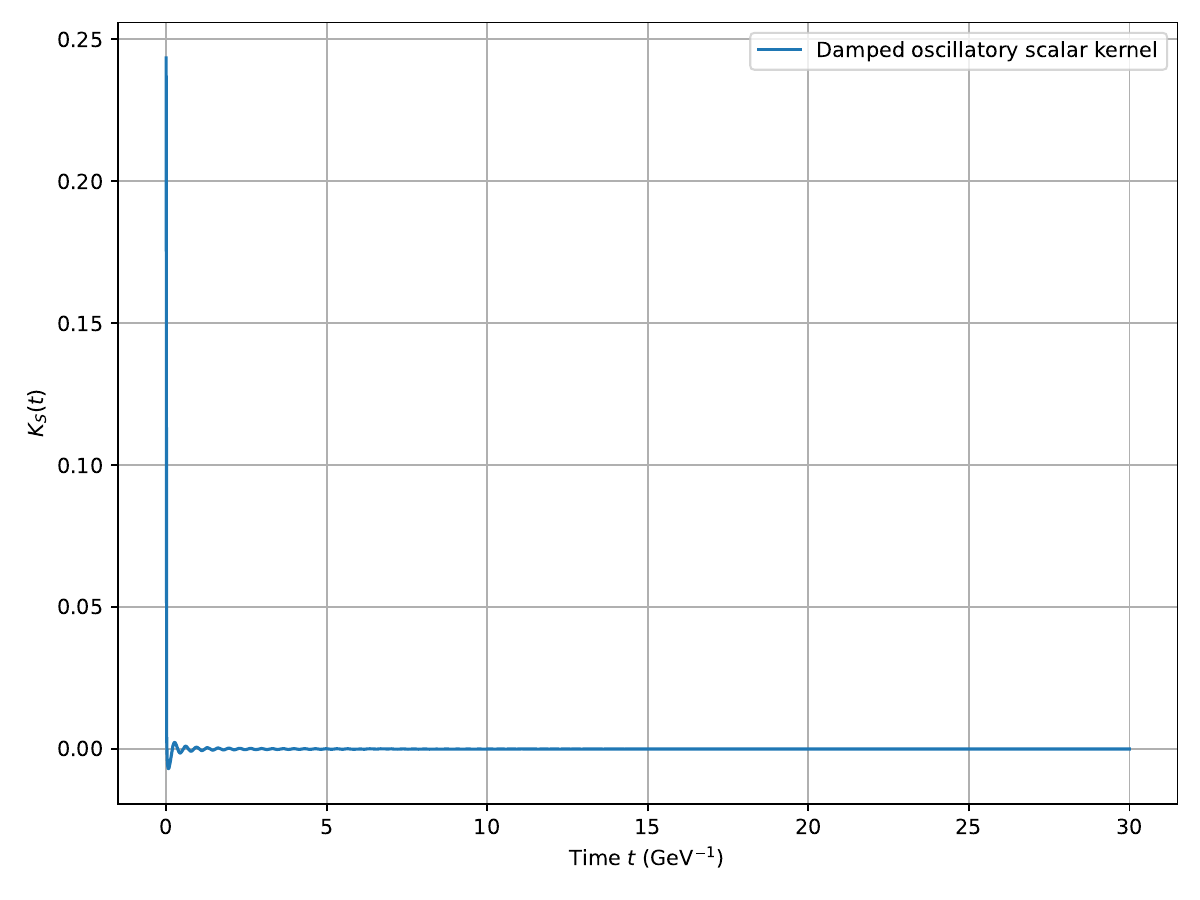}
\caption{Damped oscillatory scalar kernel $K_S(t)\propto \lambda^2 e^{-\Gamma_S t}\sin(M_S t)/(M_S t)$ for a metastable mediator. Small decay width $\Gamma_S$ produces long-lived oscillations that can source echo-like features in the GW spectrum.}
\label{fig:scalar_kernel}
\end{figure}

\begin{figure}[h!]
\centering
\includegraphics[width=0.5\textwidth]{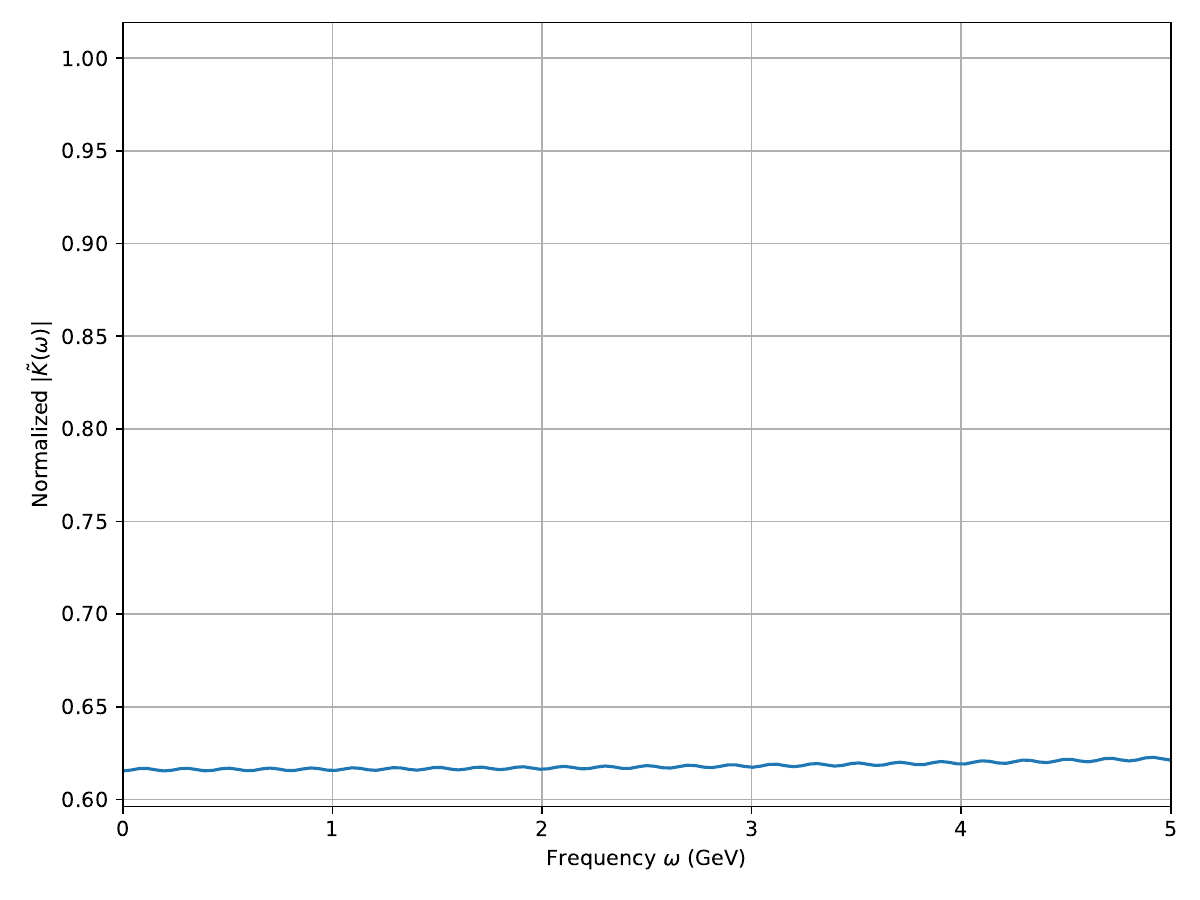}
\caption{Spectral amplitude $|\tilde K(\omega)|$ for a representative fermion-loop kernel (example $M\sim T$). The spectral shape controls how the kernel modifies the effective frequency dependence of the GW source.}
\label{fig:fermion_spectrum}
\end{figure}

Finally, Fig.~\ref{fig:fermion_spectrum} shows the spectral amplitude $|\tilde K(\omega)|$ for a representative fermion-loop kernel (here $M\sim T$), which illustrates the frequency support of the kernel and the possible overlap with GW source timescales.

\subsection{Numerical method and caveats}

Numerical integrals of Eq.~\eqref{eq:K_numeric} were computed with direct 1D quadrature in $k$ using a sufficiently large cutoff $k_{\rm max}\sim (8$–$12)\times\max(M,T)$ and uniform sampling; convergence with respect to both $k_{\rm max}$ and the number of $k$ points was checked. The integrand contains oscillatory $\sin(\omega_k t)$ factors, so the sampling density has to increase at large $t$; in practice the kernel is most relevant at microphysical times $t\lesssim \mathcal{O}(1$--$10)~{\rm GeV}^{-1}$ where our sampling is robust. The code used to generate the figures is provided in the repository and reproduces the plots in this appendix.

Two important caveats:
\begin{enumerate}
\item The fermion-loop kernel amplitude is exponentially suppressed when $M/T\gg 1$; consequently, claims of large non-Markovian effects must be supported either by choosing $M\lesssim \mathcal{O}(\text{a few})\times T$ or by invoking a meta-stable mediator with a small width.
\item The simple one-loop expressions do not capture multi-loop or collective plasma effects (e.g. Landau damping, resummed thermal masses). These effects can modify the kernel quantitatively; a full treatment requires inclusion of thermal resummation and is left to future work.
\end{enumerate}

\subsection{Implication for benchmark choices}

The numeric results clarify the parameter window relevant for the main text. If the heavy-sector masses in Table~\ref{tab:benchmarks} satisfy $M_i/T_n\gtrsim 3$ the fermion-loop contribution to the dissipative kernel is strongly suppressed (exponentially) and memory effects from such fermions are negligible unless the Yukawa couplings $y_i$ are unnaturally large. By contrast, a meta-stable scalar mediator of comparable mass with $\Gamma_S\ll M_S$ produces a long-lived oscillatory kernel and can produce observable echo features in the GW spectrum. Therefore, when claiming observable memory effects in the main text we emphasize either (i) $M_i\lesssim \mathcal{O}({\rm few})\times T_n$ or (ii) a light/metastable mediator with $\Gamma\ll M$.

\section{Modified Bounce Equation}
\label{app:bounce}

In the standard treatment of a first-order phase transition, the critical bubble profile is obtained by extremizing the three-dimensional Euclidean action,
\begin{equation}
S_3[\phi] = 4\pi \int_0^\infty dr \, r^2 \left[ \frac{1}{2} \left(\frac{d\phi}{dr}\right)^2 + V_{\rm eff}(\phi,T) \right],
\end{equation}
which yields the familiar local bounce equation,
\begin{equation}
\frac{d^2\phi}{dr^2} + \frac{2}{r}\frac{d\phi}{dr} 
= \frac{\partial V_{\rm eff}(\phi,T)}{\partial \phi}.
\label{eq:localbounce}
\end{equation}
The boundary conditions are $\phi(r \to \infty) = \phi_{\rm false}$ and $d\phi/dr|_{r=0}=0$.

\subsection{Nonlocal Effective Action}

Integrating out heavy fields with slow relaxation induces bilinear nonlocal operators in the effective action. For an $O(3)$-symmetric bounce, the action generalizes to
\begin{align}
S_3^{\rm nonloc}[\phi] &= 4\pi \int_0^\infty dr \, r^2 
\left[ \tfrac{1}{2}\left(\tfrac{d\phi}{dr}\right)^2 + V_{\rm eff}(\phi,T) \right] \nonumber \\
&\quad + 2\pi \int_0^\infty dr \, r^2 \int_0^\infty dr' \, r'^2 \, \phi(r) \, K(r-r') \, \phi(r') ,
\end{align}
where $K(r-r')$ is the spatial projection of the real-time memory kernel derived in Appendix~\ref{app:kernel}.

\subsection{Integro--Differential Bounce Equation}

Variation of this nonlocal action gives the modified bounce equation,
\begin{equation}
\frac{d^2\phi}{dr^2} + \frac{2}{r}\frac{d\phi}{dr} =
\frac{\partial V_{\rm eff}(\phi,T)}{\partial \phi}
+ \int_0^\infty dr' \, r'^2 \, K(r-r') \, \phi(r') .
\label{eq:nonlocalbounce}
\end{equation}
The last term encodes the memory effect: the field at radius $r$ couples nonlocally to its entire radial profile through $K(r-r')$.

\subsection{Kernel Models}

Two simple parametrizations are particularly instructive:
\begin{itemize}
    \item \textbf{Exponential kernel:}
    \begin{equation}
    K(r-r') \simeq \frac{1}{\tau_{\rm mem}} \, e^{-|r-r'|/\tau_{\rm mem}} ,
    \label{eq:expkernel}
    \end{equation}
    describing a heavy state with correlation time $\tau_{\rm mem}$.
    \item \textbf{Oscillatory kernel:}
    \begin{equation}
    K(r-r') \simeq \frac{\sin[M(r-r')]}{M(r-r')} \, e^{-\Gamma |r-r'|} ,
    \end{equation}
    corresponding to a meta-stable mediator with mass $M$ and width $\Gamma$.
\end{itemize}
In both cases, Eq.~\eqref{eq:nonlocalbounce} reduces to the standard bounce equation \eqref{eq:localbounce} in the short-memory limit $\tau_{\rm mem}\to 0$ or $\Gamma \to \infty$.

\subsection{Numerical Treatment}

For numerical solutions, one discretizes the radial coordinate $r$ and approximates the integral term by a quadrature sum,
\begin{equation}
\int_0^\infty dr' \, r'^2 \, K(r-r') \, \phi(r')
\;\;\longrightarrow\;\;
\sum_{j} w_j \, K(r_i-r_j)\, \phi(r_j),
\end{equation}
with $w_j$ the quadrature weights. The resulting system of coupled nonlinear equations can be solved iteratively using relaxation or shooting algorithms, in close analogy to the standard case.

\subsection{Physical Consequences}

Nonlocal contributions systematically raise $S_3/T$ at a fixed temperature, delaying nucleation and lowering the corresponding $T_n$. Bubble walls also broaden compared to the local case. Oscillatory kernels can imprint ripples in the wall profile, providing a microscopic origin for the echo features discussed in Sec.~\ref{sec:GW}. Thus, Eq.~\eqref{eq:nonlocalbounce} embeds the history dependence of the order parameter into the phase transition dynamics, directly affecting $\beta/H$, wall properties, and the resulting gravitational-wave signal.

\section{Details of GW Spectrum Modelling}
\label{app:GWspectrum}

In this appendix we summarize how nonlocal memory kernels modify the gravitational-wave (GW) templates usually employed for first-order phase transitions. The baseline decomposition separates contributions from bubble collisions (envelope approximation), long-lived sound waves, and magnetohydrodynamic (MHD) turbulence~\cite{Kosowsky:1992rz,Caprini:2007xq,Hindmarsh:2015qta,Caprini:2019egz}. 

\subsection{Standard Templates}

For reference, the standard parametrizations are
\begin{align}
\Omega_{\rm env}(f) h^2 &\simeq 1.67\times 10^{-5} \left(\frac{H_*}{\beta}\right)^2
\left(\frac{\kappa_\phi \alpha}{1+\alpha}\right)^2
\left(\frac{100}{g_*}\right)^{1/3}
S_{\rm env}(f/f_{\rm env}), \\
\Omega_{\rm sw}(f) h^2 &\simeq 2.65\times 10^{-6} \left(\frac{H_*}{\beta}\right)
\left(\frac{\kappa_v \alpha}{1+\alpha}\right)^2 v_w
\left(\frac{100}{g_*}\right)^{1/3}
S_{\rm sw}(f/f_{\rm sw}), \\
\Omega_{\rm turb}(f) h^2 &\simeq 3.35\times 10^{-4} \left(\frac{H_*}{\beta}\right)
\left(\frac{\kappa_{\rm turb} \alpha}{1+\alpha}\right)^{3/2}
\left(\frac{100}{g_*}\right)^{1/3}
S_{\rm turb}(f/f_{\rm turb}),
\end{align}
with $H_*$ the Hubble rate at the transition, $\beta$ its inverse duration, $\alpha$ the strength parameter, $g_*$ the effective relativistic degrees of freedom, and $\kappa_i$ the efficiency factors for converting latent heat into each channel. The spectral functions $S_i$ describe the shape of the signal around its characteristic frequency $f_i$.

\subsection{Smearing by Memory Effects}

With a finite memory time $\tau_{\rm mem}$, the latent heat release is not instantaneous. Instead, the stress-energy tensor sourcing the GW equation,
\begin{equation}
\ddot{h}_{ij}(t,\mathbf{k}) + 3H\dot{h}_{ij}(t,\mathbf{k}) + k^2 h_{ij}(t,\mathbf{k})
= 16\pi G \, \Lambda_{ij,kl} T^{kl}(t,\mathbf{k}),
\end{equation}
is replaced by a temporally smeared version
\begin{equation}
T^{kl}_{\rm eff}(t) = \int^t dt' \, K(t-t') \, T^{kl}(t').
\end{equation}
In Fourier space, this corresponds to
\begin{equation}
\tilde{T}^{kl}_{\rm eff}(\omega) = \tilde{K}(\omega)\,\tilde{T}^{kl}(\omega).
\end{equation}
Thus the spectral shape functions are modified as
\begin{equation}
S_i(f/f_i) \;\longrightarrow\; \tilde{K}(2\pi f)\, S_i(f/f_i).
\end{equation}
For example, an exponential kernel gives $\tilde{K}(\omega)=1/(1+i\omega\tau_{\rm mem})$, producing Lorentzian broadening of the peaks.

\subsection{Echo Contributions}

If the kernel contains oscillatory components,
\begin{equation}
K(t) \sim e^{-\Gamma t}\cos(M t),
\end{equation}
its Fourier transform develops sidebands near $\omega=\pm M$. This generates secondary contributions to the GW spectrum, which can be schematically represented as
\begin{equation}
\Omega_{\rm GW}(f) \simeq \Omega_{\rm base}(f) 
+ \sum_{n=1}^{N_{\rm echo}} c_n\, e^{-\gamma n}\,
\Omega_{\rm base}(f+n\Delta f_{\rm echo}),
\end{equation}
with echo spacing
\begin{equation} \label{eq:fecho}
\Delta f_{\rm echo} \simeq \frac{M}{2\pi a_0}
\left(\frac{T_0}{T_*}\right)\left(\frac{g_0}{g_*}\right)^{1/3},
\end{equation}
where $a_0$ and $T_0$ are today’s scale factor and temperature. The coefficients $c_n$ capture the (model-dependent) amplitude of each echo. While the precise strength is sensitive to the microscopic structure of the kernel, the appearance of sidebands is a generic feature of oscillatory nonlocalities.

\subsection{Implementation in Benchmarks}

For the benchmarks BP1--BP3 in the main text, we adopt a practical scheme:  
\begin{itemize}
\item multiply each standard template by $\tilde{K}(\omega)$ to account for broadening,  
\item add echo contributions when the kernel contains oscillatory terms, using the parametric form above.  
\end{itemize}
This procedure interpolates between the standard case (BP1), broadened spectra (BP2), and spectra with visible echoes (BP3). The latter case corresponds to long-lived oscillatory kernels where the sidebands can lie in the mHz–Hz window. 

In summary, memory kernels alter the GW spectrum in two characteristic ways:
\begin{enumerate}
  \item \textbf{Smearing:} finite $\tau_{\rm mem}$ broadens peaks and suppresses amplitudes, similar to a Lorentzian filter.  
  \item \textbf{Echoes:} oscillatory kernels induce secondary structures with spacing controlled by Eq.~\eqref{eq:fecho}.  
\end{enumerate}
The echo effect should be regarded as a well-motivated possibility rather than a guaranteed outcome, with its detectability dependent on whether the oscillatory component of $K(t)$ survives plasma damping. Within these caveats, the resulting template modifications are straightforward to implement and provide distinctive targets for future detectors.

\bibliographystyle{apsrev4-2}
\bibliography{memory_U1X}

\end{document}